\documentclass[twocolumn,superscriptaddress,secnumarabic,amssymb,nobibnotes,nofootinbib,aps,prl,showpacs]{revtex4}
\usepackage{graphicx}
\begin{document}
\setlength{\topmargin}{-0.05in}

\title{
 Model study on the photoassociation of a pair of
trapped atoms \\
into an ultralong-range molecule}

\author{B. Deb}
 \affiliation{ Physical
Research Laboratory, Navrangpura, Ahmedabad 380 009, India}

\author{ L. You}
\affiliation{School of Physics, Georgia Institute of Technology,
Atlanta, Georgia 30332, USA}

\date{\today}

\def\zbf#1{{\bf {#1}}}
\def\bfm#1{\mbox{\boldmath $#1$}}
\def\hf{\frac{1}{2}}

\begin{abstract}
Using the method of quantum-defect theory, we calculate the
ultralong-range molecular vibrational states near the dissociation
threshold of a diatomic molecular potential which asymptotically
varies as $-1/R^3$. The properties of these states are of
considerable interest as they can be formed by photoassociation
(PA) of two ground state atoms. The Franck-Condon  overlap
integrals between the harmonically trapped atom-pair states and
the ultralong-range molecular vibrational states are estimated and
compared with their values for a pair of untrapped free atoms in
the low-energy scattering state. We find that the binding between
a pair of ground-state atoms by a harmonic trap has significant
effect on the Franck-Condon integrals and thus can be used to
influence PA. Trap-induced binding between two ground-state atoms
may facilitate coherent PA dynamics between the two atoms and the
photoassociated diatomic molecule.
\end{abstract}


\maketitle

\section{introduction}
Photoassociation (PA) spectroscopy \cite{revpa} has developed into
an important tool for studying properties of cold atoms and
diatomic molecules. The formation of excited diatomic molecular
states through cold atom PA had been reported a long time ago
\cite{thorsheim}. Over the last decade, theoretical techniques for
calculating PA rate coefficients have been developed by many
groups \cite{theory}. Photoassociation spectroscopy enables a
precise calibration of long-range interatomic forces. It leads to
the formation of translationally cold molecules and thus provides
one possible route of creating a molecular Bose-Einstein
condensate (BEC). Two-photon resonant Raman PA coupling can also
be utilized for generating atom-molecule coherence
\cite{drummond1,javanainen1} and massive entangled atoms
\cite{li}. In fact, many-body atom-molecule coherence was recently
observed \cite{wieman} through the tuning of a Feshbach resonance
(FR) \cite{feshbach}  caused by an applied magnetic field in a
$^{85}$Rb condensate. Quasibound molecular states can be formed by
FR in ultracold atom-atom collision. FR has striking analogy with
PA. In the first approximation, both are coupled two-state systems
and mathematically identical. It is the magnetic field which
couples the two states  in the former case, while in the latter
case it is the optical field. Both can be used to change the
scattering length  between two atoms \cite{fatemi}. Two-color
stimulated Raman adiabatic passage (STIRAP)
\cite{javanainen2,drummond2,quasicont,hope} via coherent PA using
an appropriately tailored pair of laser pulses has been proposed
for creating a molecular condensate from an atomic BEC. Stimulated
PA of condensed atoms leads to ``superchemistry"
\cite{superchem,supchem} where quantum statistics plays a crucial
role in the production of molecules, and a superposition of
macroscopic number of atoms and molecules \cite{atom-molecule}.
Although there has been no explicit experimental realization of a
molecular BEC, coherent production of molecules in a single
rotational-vibrational state has recently been experimentally
demonstrated \cite{becmolecule}. The quantum dynamics \cite{qdyn}
of coupled atomic and molecular condensates reveals a lot of
interesting physical effects, e.g., the recently observed
frequency shifts \cite{shifts} contains the detailed information
about the density of states of cold atoms in the quasicontinuum
regime. An alternative approach \cite{mbecmott} for creating a
molecular BEC from atoms in a Mott insulating state inside an
optical lattice was also proposed recently \cite{mott}.

In a PA event, two ground-state atoms, by means of a single-photon
excitation during the collision, combine to form an excited
diatomic molecule which can decay back to two atoms or to a
different ground molecular state. Near-zero-energy vibrational
states close to the dissociation threshold of an excited diatomic
molecular level can be formed by PA of cold atoms. These states
can extend to several hundred nanometers. PA absorption line
strength and linewidth strongly depend on the overlap integral,
known as Franck-Condon (FC) integral, of these states with two
atoms in the ground state. For these ultralong-range states, the
``reflection approximation" which is often used in molecular
spectrum calculations, breaks down. The reflection approximation
replaces the actual integral by an integral over a $\delta$
function. This drastically simplifies the calculation by ruling
out the necessity of calculating the wave function in the entire
range. Instead, it only requires to calculate the wave function at
a particular interatomic separation (the turning, or Condon
point). This indeed seems to be a valid approximation for deeply
bound molecular states with energy several tens of GHz below the
dissociation threshold. For bound states with energies of the
order of MHz or even kHz below the dissociation threshold,
however, this approximation does not seem to work. Calculation of
the actual Franck-Condon overlap integral is, therefore,
inevitable for such states. This presents a significant challenge
as the standard numerical techniques for calculating the wave
functions of these states are not known to be reliable.

In this paper, we discuss a scheme for estimating the qualitative
dependence of the required FC integrals. By using the
quantum-defect theory (QDT) to find the exact solutions of
molecular states in a long-range excited potential ($-1/R^3$), we
obtain near-zero energy ($\le$ MHz) vibrational wave functions
that can extend from several hundred nanometers  to micrometers.
We employ these wave functions to evaluate the FC overlap integral
in two situations: (1) for two atoms trapped in a harmonic
potential and (2) for two free atoms. We analyze the effect of the
trapping potential and the atom scattering length on the FC
integral. We also evaluate spontaneous emission linewidths of the
excited molecular states.

This paper is organized as follows. In the following section, we
describe the model system and the basis for our study. Then in
Sec. III, we discuss the relative motional state of two
ground-state (S+S) atoms trapped in a harmonic potential.  In Sec.
IV, we describe the method of QDT for calculating the vibrational
states that can be formed by PA in the excited electronic
potential which asymptotically corresponds to two (S+P) separated
atoms. Photoassociation of two atoms inside an isotropic harmonic
trap, with an emphasis on the usefulness of the trap-induced
two-atom bound states in one- and two-color coherent PA, is
discussed in Sec. V. We present and discuss the results of our
study in Sec. VI. We conclude in Sec. VII.

\section{the model system}
Using the Born-Oppenheimer approximation, the Hamiltonian of our
model with two atoms $A$ and $B$ can be expressed as $H =
H_{\mathrm nucl} + H_{\mathrm el} + V_{\mathrm trap} + V_{\mathrm
af}$,
 where
\begin{eqnarray}
H_{\mathrm nucl} &=& -\frac{\hbar^2}{2M}\nabla_{\rm CM}^2 -
\frac{\hbar^2}{2\mu}\nabla_{\mathrm rel}^2 + V(R).
\end{eqnarray}
$M$ and $\mu$ are the total and the reduced mass of the two atoms,
respectively. The Laplacians $\nabla_{\mathrm CM}^2$ and
$\nabla_{\mathrm rel}^2$, respectively, correspond to the center
of mass and the relative nuclear coordinate, of the two atoms.
$V(R)$ is the atom-atom interaction potential which is
approximated as isotropic, i.e., as a function of the relative
nuclear coordinate $R = |{\mathbf{R}}_A - {\mathbf{R}}_B|$ only.
$H_{\rm el}$ is the electronic part of the total Hamiltonian and
$V_{\rm trap} = \sum_{\alpha=A,B} V_{\rm trap}^{\alpha}$ with
$V_{\rm trap}^{\alpha}$ as the trapping potential for atom
$\alpha$. The atom-field interaction is given by the standard
dipole approximation
\begin{eqnarray}
V_{\rm af} = -\sum_{\alpha=A,B}
{\mathbf{E}}\cdot{\mathbf{d}}_{\alpha}, \label{dp}
\end{eqnarray}
where ${\mathbf{E}}$ represents the electric field of the applied
laser field and ${\mathbf{d}}_{\alpha}$ denotes the electronic
dipole moment of atom $\alpha$. In the limit of a low-intensity
laser, this atom-field interaction can be considered as a
perturbation.

We choose molecular states as our basis functions. Initially, the
two atoms $A$ and $B$ are in their ground $s$ electronic states.
Thus their initial molecular state notation is $|\Phi_{i} \rangle
= |ns,ns,^{2S+1}\!\Lambda_{g,u},\epsilon_{i},J_i,M_i\rangle
|\Psi_{i}(R)\rangle$, where $\Lambda$ is the projection of the
total electronic orbital angular momentum on the molecular axis
and $\epsilon_{i}$ is the relative energy of the two atoms.
$|\Psi_i(R)\rangle$ is the initial relative nuclear wave function.
The final molecular state can be represented by $|\Phi_{f} \rangle
= |ns,np,\Omega_{u,g}, \epsilon_{v},J_f,M_f\rangle
|\Psi_{v}(R)\rangle$, where $v$ denotes the vibrational quantum
number, the energy $E_v$ of the vibrational state $v$ lies at an
energy $\epsilon_{v} = D_f - E_v$ below the dissociation threshold
$D_f$. Here $\Omega_{u,g}$ refers to  Hund's case c of the
interatomic potential; $J_{i(f)}$ is the total angular momentum
(consisting of electronic orbital, spin, and molecular orbital
angular momenta) of the initial (final) state with respect to a
laboratory axis and $M_{i(f)}$ is its projection on that axis.
$|\Psi_v(R)\rangle$ is the relative vibrational nuclear wave
function. These basis states are similar to those used in Ref.
\cite{pillet}. In this representation of the basis states, we have
neglected the hyperfine interaction for the sake of simplicity.
The dipole interaction term (\ref{dp}) in this molecular basis, in
the center-of-mass frame of the two atoms, then takes the form
\begin{eqnarray}
 V_{\rm af} \simeq
-\hat{\epsilon}\cdot\hat{\mu} E
D(R)\cos({\mathbf{k}}_{L}\cdot{\mathbf{R}}/2), \label{vaf}
\end{eqnarray}
where $D(R)$ is the molecular dipole matrix element between the
molecular states $\Lambda_{g,u}$ and $\Omega_{u,g}$,
${\mathbf{k}}_{L}$ and $\hat{\epsilon}$ are the wave vector and
the polarization unit vector, respectively, of the laser field;
and  $\hat{\mu}$ is the unit vector of the dipole moment. For two
homonuclear alkali metal atoms, $D(R)$ has been calculated by
Marinescu and Dalgarno \cite{marinescu}. To leading order, it
asymptotically varies as $1/R^3$ in the form of $D(R) = D_0 +
D_1/R^3$. For the $3S_{1/2}-3P_{3/2}$ transition in Na$_2$, in
atomic units $D_0=-3.5007$ and $D_1 = 142.13$ \cite{marinescu}.
Therefore, for $R$ larger than 100${a_0}$ in unit of Bohr radius
$a_0$, the $R$ dependence of $D(R)$ of Na$_2$ can safely be
neglected and $D(R)$ can be approximated as a constant $D_0$. All
ultralong-range excited vibrational states of Na$_2$ we report
here have negligible amplitude in $R<100$; instead, their major
amplitude lies at a separation much greater than a 1000 Bohr
radii.

To illustrate our study with concrete results, we specifically
consider the transition $^3\Sigma_{u} \rightarrow 0_{g}^{-}$ of
Na$_2$. The initial and the final internal states are assumed to
be $|ns,ns,^3\!\Sigma_u,\epsilon_{i},0,0\rangle$ and
$|ns,np,0_g^{-},\epsilon_{v},1,0\rangle$, respectively. The
potential of the $0_g^{-}$ and $1_u$ states of Na$_2$ has been
explicitly tabulated by Stwalley {\it et al.} \cite{stwalley}.
These states have their equilibrium positions at a separation
larger than 60$a_0$. Compared to the size of a ground diatomic
molecule whose equilibrium position ranges in general between 2
and 10 Bohr radius, these $0_g^{-}$ and $1_u$ molecular states are
truly long ranged. The PA spectrum for some vibrational states of
$0_g^-$ of Na$_2$ have already been observed experimentally
\cite{0g}.

\section{two-atom state in a trap}
We can choose the initial state of two ground-state (S) atoms in
two different ways depending on whether the two atoms are trapped
or are moving freely. The corresponding PA spectrum for these two
cases can become markedly different. For two free ground-state
atoms colliding under the interatomic interaction, their
asymptotic relative motion scattering wave function is given by
\begin{eqnarray}
\Psi_{\epsilon} = \sqrt{\frac{k}{\pi\epsilon}}\sin[k(R-a_{\rm sc})]
\end{eqnarray}
which is an energy-normalized scattering state. $\epsilon
=\hbar^2k^2/(2\mu)$ is the asymptotic collision energy. At low
energies, for $|R-a_{\rm sc}| \ll k^{-1}$ this wave function can
be approximated as a straight line. Care should be taken to
calculate the free-bound FC factor of ultralong-range molecular
states, since the range of the vibrational state can exceed
$k^{-1}$ even at very low energies.

For two atoms in an isotropic harmonic trap, their wave function
is separable into the center of mass and the relative motion
\cite{wilkens}.  Approximating the atom-atom  interaction $V(R)$
by a regularized contact potential $4\pi \hbar^2 a_{\rm
sc}\delta^{3}({\mathbf R})/(2\mu)$ (with $a_{\rm sc}$ as the
s-wave scattering length), the bound state of two atoms in a
harmonic trap (of frequency $\omega_t$) was first derived
analytically by Busch {\it et al.} \cite{wilkens}.  Let us define
a characteristic length scale of an isotropic  harmonic trap as
$a_{t} = \sqrt{\hbar/(\mu\omega_t})$ and introduce a dimensionless
quantity $\bar{R} = R/a_t$. Let $\Psi_{n_t} = R\Phi_{n_t}$
represent the relative motional bound-state wave function in the
harmonic trap, with $n_t$ (which we will define below) being an
integer quantum number related to the three-dimensional isotropic
harmonic trap. We can then express the bound state as
\begin{eqnarray}
\Phi_{n_t}  =
\frac{1}{2}\pi^{-3/2}A\exp(-\bar{R}^2/2)\Gamma(-\nu)U(-\nu,\frac{3}{2},
\bar{R}^2),
\end{eqnarray}
where $A$ is a normalization constant having the dimension of
inverse of square root of volume   and
\begin{eqnarray}
\nu = \frac{1}{2\hbar\omega_{t}}\epsilon- \frac{3}{4},
\end{eqnarray}
an effective quantum number  for the relative motional eigenstate.
$U(.,.,.)$ is the confluent hypergeometric function. The energy
spectrum for such trap-bound two-atom states has been analyzed in
detail before \cite{wilkens}. The energy eigenvalue is given by
the roots of the equation
\begin{eqnarray}
\frac{\Gamma(-x/2 + 3/4)}{\Gamma(-x/2+1/4)} =
\frac{1}{\sqrt{2}\xi_s},
\end{eqnarray}
where $x = \epsilon/(\hbar\omega_t)$.
\begin{table}
\caption{The energy eigenvalues  $E_{\nu}$ of the lowest three
harmonic trap-bound  s-wave vibrational atom-pair states and their
outer turning point $R_t$ for two values of harmonic frequencies.}
\begin{tabular}{c c c c c}
 &\multicolumn{2}{c}{$\omega_t = 2\pi \times 10$ kHz}
 &\multicolumn{2}{c}{$\omega_t = 2\pi \times 100$ kHz} \\
 \multicolumn{1}{c}{$n_t(\nu)$} &\multicolumn{1}{c}{$E_{\nu}$ (MHz)}
 &\multicolumn{1}{c}{$R_t$ (nm)}
  &\multicolumn{1}{c}{$E_{\nu}$ (MHz)}
  &\multicolumn{1}{c}{$R_t$ (nm)}\\
  \hline
0 & 0.095 & 515.44 & 0.96 &164.25 \\
1 & 0.221 & 786.35 & 2.23 & 249.89 \\
2 & 0.347 & 985.28 & 3.49 & 312.78
  \end{tabular}
\label{tb1}
\end{table}
 We here define a
dimensionless quantity $\xi_s = a_{\rm sc}/a_t$. If $\xi_s \ll 1$,
then the eigenenergies can be approximated as
\begin{eqnarray}
\epsilon_{n_t} \simeq \left[\frac{3}{2} + 2n_t + \sqrt{2/\pi}\xi_s
C_{n_t}^{n_t+1/2}\right]\hbar\omega_t, \hspace{0.2cm}
n_t=0,1,2,....
\end{eqnarray}
\begin{figure}
\includegraphics[width=3.25in]{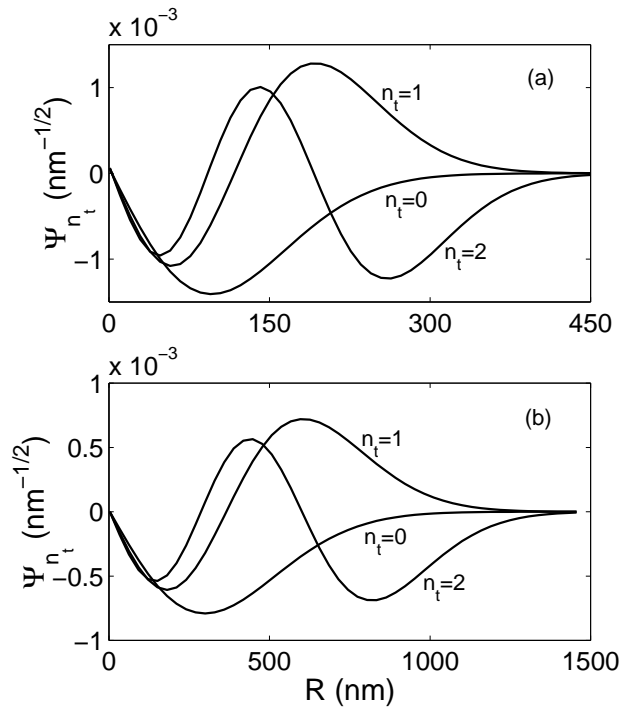}
 \caption{Two $^{23}$Na atom s-wave
bound-state wave function (in units of nm$^{-1/2}$) in a harmonic
trap with $\omega_{t} = 2\pi \times 100$ (kHz) (a) and $\omega_{t}
= 2\pi \times 10$ (kHz) (b).
 The individual curves are specified by the trap quantum
 number $n_t$.}
\label{fig1}
\end{figure}

Selected numerical results are presented in Table \ref{tb1} and
Fig. \ref{fig1}. Recently, the energy spectrum of two atoms in a
harmonic trap has also been analyzed in detail in two papers
\cite{blume-bolda} where it was shown that the validity of the
solutions found in Ref.\cite{wilkens} critically depends on a
characteristic length scale of the interatomic potential. This
length scale is defined by $\beta_6 = (2\mu C_6/\hbar^2)^{1/4}$,
where $C_6$ is  van der Waal's coefficient and $\mu = m/2$ is the
reduced mass of the two atoms. If both $\beta_6 \ll a_{t}$ and
$a_{\rm sc} \ll a_t$, then the solutions of Ref. \cite{wilkens}
are physically valid. If these validity conditions are not
satisfied, then the use of an energy-dependent scattering length
in the pseudopotential improves the solution as demonstrated in
Ref.\cite{blume-bolda}. For a trap frequency $\omega_t = 2\pi
\times 100$ kHz and for $^{23}$Na atoms, we find
 $\xi_s = a_{\rm sc}/a_t = 0.042$
and $\beta_6/a_t = 0.13$. Thus for our model system pursued in
this study as described in Sec. II., we can safely use the
solutions of Ref.\cite{wilkens} for very strong traps with
frequency $\omega_t \le 2\pi \times 100$ (kHz).

\section{Bound states in the $-C_{3}/R^3$ potential}

In order to compute the PA spectrum, one first needs to find the
vibrational wave functions of the excited diatomic molecule. It
turns out, however, that it is not always necessary to find
vibrational states for the entire range--the knowledge of wave
functions near the vicinity of an outer turning point sometimes
allows one to make a good estimate of the FC factor. In such
cases, the FC factor can be approximated as the product of the two
vibrational amplitudes at the outer turning point. This
approximation, known as reflection approximation, relies on the
fact that the molecule has its major probability of existence at
the semiclassical outer turning point. Barring the ground state,
this is true for most of the low-lying vibrational states of a
molecule. But, for the vibrational states near dissociation
threshold, the probability can be distributed over a wide range.
Therefore, for those states, the validity of this approximation
becomes questionable. This was examined in the past by many
authors for the case of bound-free transitions \cite{bound-free}
that defy the reflection approximation.

In this study, we concentrate on near-threshold vibrational bound
states in a potential that varies asymptotically as $-1/R^3$,
which corresponds to a S+P excited molecular state between two
atoms in the separated atom limit. Recently, these states have
become very important in view of their wide accessibility in
various PA processes. It is widely appreciated that there exists a
class of pure long-range excited molecules whose equilibrium
position is at a large separation \cite{stwalley}. The binding
energies of these vibrational states can range from a few (Hz) to
several hundred GHz. The lower the binding energy is, the longer
the spatial extension of the states is. In calculating the FC
factor for these states with respect to the state of two ground
state atoms moving either freely or in a trap, use of the
reflection approximation becomes questionable, particularly so for
the vibrational states with sub-GHz energies. Therefore, it is
necessary to calculate the vibrational states in the entire range
or at least in a broader range near the outer turning point.

In practice, it is difficult to accurately compute wave functions
for vibrational states just below the dissociation threshold by
the standard numerical integration procedure such as the Numerov
method. Therefore, alternative procedures such as the
semi-analytic approach offered by the method of QDT need to be
explored. While QDT has been successfully employed over the years
for computing Rydberg-atom-like energy spectrum, its application
to the problem of PA has not yet been fully explored. For
Rydberg-atom-like systems, the relevant potential is the Coulomb
potential that admits analytic solutions known as Coulomb
functions which forms the basis for the application of QDT. For
the excited state in the single-photon PA process as discussed
here, the potential is of an attractive $-1/R^3$ type.
Fortunately, a semianalytic QDT \cite{bohn} of PA has recently
been developed which is based on the semiclassical WKB-type
solutions of a $-1/R^3$ potential. Exact quantum solutions of this
potential have also been found recently \cite{bogao,deb}. For some
diatomic molecules, the spacings of near zero-energy vibrational
states can be as small as a few kHz or even a few Hz, which leads
to the interesting possibility that the free-bound PA linewidth of
these states can exceed the spacings by several times even at a
temperature as low as 1 $\mu$K. Such a situation prevents the
resolution of individual spectral lines. One naive suggestion to
overcome this problem is to place two atoms inside a single trap,
and to photoassociate them for different trap frequencies.
Although PA in a trap may lead to a diffused continuous absorption
spectrum, as we will show below, the use of different trap
frequencies results in shifts of the peak of the spectrum. The
energies of the different vibrational levels can be estimated from
these shifts.

We now  briefly outline the use of QDT for computing
near-threshold ultralong-range vibrational states. The exact
scattering solution of a repulsive $1/R^3$ was first obtained by
Gao \cite{bogaor3}, and the exact analytic solutions of an
attractive $1/R^3$ have been applied to a QDT formulation of
diatomic molecular excited vibrational energy calculations before
\cite{bogao}. The scattering solutions of an attractive $1/R^3$
potential has also been used before for a multichannel QDT
formulation of atom-atom interactions in the presence of a dc
electric field \cite{deb}.

We next concentrate on the bound
solutions of a $-1/R^3$ potential. The asymptotic form of linearly
independent base functions $f_{l}$ and $g_{l}$ satisfying the
Schr\"odinger equation with this potential for energy $E<0$ can be
expressed as
\begin{eqnarray}
 \left( \begin{array}{cc} f_{l}(R\rightarrow \infty)
\\ g_{l}(R\rightarrow \infty) \end{array} \right)
=\left( \begin{array}{cc} W_{f-} & W_{f+}
\\ W_{g-} & W_{g+} \end{array} \right)
 \left( \begin{array}{cc} \exp(\kappa R)
\\ \exp(-\kappa R) \end{array} \right),
\end{eqnarray}
where $W$'s are chosen to be real functions
\begin{eqnarray}
W_{f\mp} &=&\frac{\exp(\pm
i\theta)}{D(\nu)}\sqrt{\frac{2}{\pi\kappa}}\left[\frac{C_{+}(\nu)}{G(-\nu)}\xi_{\mp}
-\frac{C_{+}(-\nu)}{G(\nu)}\eta_{\mp}\right],\nonumber\\
W_{g\mp}&=&\frac{\exp(\pm
i\theta)}{D(\nu)}\sqrt{\frac{2}{\pi\kappa}}
\left[-\frac{C_{-}(\nu)}{G(-\nu)}\xi_{\mp}+
\frac{C_{-}(-\nu)}{G(\nu)}\eta_{\mp}\right],
\end{eqnarray}
with $\theta = \pi\nu/2 + \pi/4$ and
\begin{eqnarray}
\xi_{\mp} &=& \sum_{m=-\infty}^{\infty} b_m  \exp(\pm im\pi/2),\\
\eta_{\mp} &=& \sum_{m=-\infty}^{\infty} (-1)^m b_m  \exp(\mp
im\pi/2).
\end{eqnarray}
Following standard conventions, we take
$\hbar^2\kappa^2/(2\mu)=-E$, the bound state energy. The other
parameters $C_{\pm}(\nu)$, $D(\nu)$, $G(\nu)$ and $b_m$  are as
defined in  appendix A of Ref.\cite{deb}. The condition for bound
states is given by
\begin{eqnarray}
\chi - K^{0}=0, \label{cbeq}
\end{eqnarray}
where $K^{0}$ is the short-range K matrix determined by matching
the solutions of short- and long-range potentials at a judiciously
chosen separation, and $\chi$ is given by
\begin{eqnarray}
\chi=-W_{f-}/W_{g-}.
\end{eqnarray}
With the help of QDT, the short range K matrix is extrapolated
from the positive energy scattering solutions near the
dissociation threshold. The bound-state wave function at discrete
energy $E_{i}$ can then be expressed as
\begin{eqnarray}
\Psi_{i}(R)=f_{l}- K^{0}g_{l}. \label{bs}
\end{eqnarray}

\begin{figure}
\includegraphics[width=3.25in]{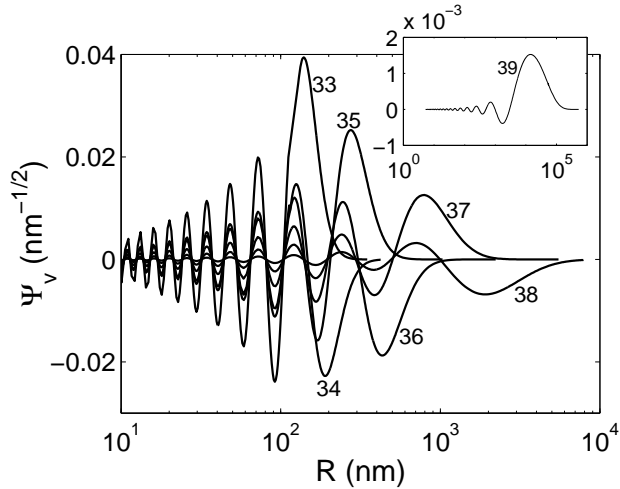}
 \caption{The last 7
(from 33 to 39) s-wave vibrational wave functions (in nm$^{-1/2}$)
 of the $0_g^{-}$ state of Na$_2$. The numbers assigned to the
 curves refer to the
 vibrational quantum numbers. The range of the
 last bound state ($v=39$)  shown in the
 inset is so long that it exceeds $\mu$m by two orders of magnitude.}
\label{fig2}
\end{figure}

Selected numerical results for the ultralong range molecular
states are given in Table \ref{tb2} and Fig.\ref{fig2}.

\begin{table}
\caption{The energy $E_v$ of the last seven molecular vibrational
levels of the $0_g^-$ state of a Na$_2$ for rotational quantum
number $l=0$ and their outer turning point $R_t$ and the
separation $R_{\rm max}$ at which the probability $|\Psi_v(R)|^2$
peaks.}
\begin{tabular}{c c c c }
\multicolumn{1}{c}{$v$} &\multicolumn{1}{c}{$E_v$ (kHz)}
 &\multicolumn{1}{c}{$R_t$ (nm)}
&\multicolumn{1}{c}{$R_{\rm max}$ (nm)}\\
\hline
 33 & 1460 & 162.3 & 139.3    \\
 34 & 550 & 224.8 & 190.2   \\
 35 & 170 & 332.3 & 275.4   \\
 36 & 39.5 & 540.3 & 431.5  \\
 37 & 5.7 & 1000 & 794.0   \\
 38 & 0.3017 & 2700 &  1900    \\
 39 & 0.0003 & 28200 & 14200
\end{tabular}
\label{tb2}
\end{table}

\section{photoassociation in a harmonic trap}
A key factor of the PA spectrum strength is the FC integral. Ever
since its introduction more than 70 years ago
\cite{franck-condon}, the FC principle remains a paradigm in
molecular spectroscopy, particularly for the spectra of diatomic
molecules. The essence for this principle relies on the wave
mechanical property of molecules. The underlying idea is that when
a molecule undergoes electronic transition  from one state to
another, the transition occurs much faster than the time scale of
the vibrational motion of its constituent atomic nuclei.
Therefore, just before and after the transition, the relative
nuclear position remains almost the same. The wave mechanics of
the molecule comes into the picture when one calculates the
transition probability between the two states. In view of the FC
principle, the strength of the transition is determined by an
integral over the overlap of initial and final wave functions of
the nuclear motion. The line strength and the linewidth of
molecular spectrum are proportional to this overlap integral. The
ground vibrational state of a molecule has the maximum probability
at its equilibrium position, while the excited vibrational states
have their maximum probability near outer turning points.
Accordingly, the relative location of the equilibrium positions is
also important in determining which vibrational transition would
have a given intensity.

For two atoms initially in a harmonic trap, we can define a Rabi
frequency
\begin{eqnarray}
\Omega_{n_t-v} = \frac{1}{\hbar} |\langle n_t | V_{\rm af} | v
\rangle| \label{rb}
\end{eqnarray}
between the trap-induced bound state  $|n_t \rangle$ of the two
atoms and the excited molecular vibrational state $|v \rangle$ due
to the interaction of the two-atom system with the PA laser field.
The Rabi frequency Eq. (\ref{rb}) is proportional to the rate of
coherent population transfer  between the two-atom state $|n_t
\rangle$ and the  molecular state $|v \rangle$. Clearly, it is
proportional to overlap integral
\begin{eqnarray}
\eta_{v-n_t} = \int \Psi_v\Psi_{n_t}\cos({\mathbf
k}_L\cdot{\mathbf R}) dR,
\end{eqnarray}
which is the FC factor between the two states. Since both states
$|v\rangle$ and $|n_t\rangle$ are discrete (bound) and unit
normalized, the FC factor in this case is dimensionless.

Here we wish to stress that the dynamics of photoassociation of
two trapped atoms is strikingly different from that of  two free
atoms. Since two trapped atoms form discrete relative motional
states, there exists a well-defined Rabi frequency and the
associated coherent dynamics. In contrast, for two freely moving
atoms, the definition of free-bound Rabi frequency is not so
straightforward, since Rabi frequency is usually defined between
two discrete states. Instead, for low laser power, applying
Fermi's golden rule, one can define free-bound stimulated
transition rate
\begin{eqnarray}
\Gamma_{{\rm st}}^{v-\epsilon} = \frac{2\pi}{\hbar} |\langle v
|V_{\rm af} | \epsilon \rangle |^2,
\end{eqnarray}
which is proportional to the square of the free-bound FC factor
\begin{eqnarray}
\eta_{v-\epsilon} = \int \Psi_v\Psi_{\epsilon}\cos({\mathbf
k}_L\cdot{\mathbf R})dR, \label{fb}
\end{eqnarray}
having the dimension of $1/\sqrt{\epsilon}$, i.e.,
$|\eta_{v-\epsilon}|^2$ has dimension of per unit energy.
$\Gamma_{{\rm st}}^{v-\epsilon}$  is the free-bound stimulated
decay rate and does not characterize any coherent PA. In order to
describe free-bound coherent PA dynamics, a quasicontinuum model
\cite{quasicont} has been proposed. In this model, quasi-continuum
is assumed to consist of many discrete equally spaced levels and
thus enables to define a photoassociative Rabi frequency. Then the
continuum limit is taken by allowing the energy spacing between
the levels to approach zero. With this model analysis, it was
shown \cite{quasicont} that the free-bound Rabi frequency is
proportional to the square root of the collision energy. Thus in
the zero energy limit, the free-bound Rabi frequency vanishes and
so there is no coherent photoassociative dynamics. Alternatively,
the quasicontinuum model can be supplemented \cite{javanainen2} by
introducing two-atom scattering states normalized within a sphere
of arbitrary volume ${\cal V}$.  After all the calculations are
done, the continuum limit can be taken by allowing the volume
${\cal V}$ to go to infinity. It is thus argued that for a
nondegenerate gas, free-bound Rabi frequency is much smaller than
the corresponding bound-bound Rabi frequency. To transfer atoms
into molecules by STIRAP, one of the necessary and sufficient
conditions is the swapping of the bound-bound Rabi frequency into
free-bound Rabi frequency in a counterintuitive way, i.e., for an
initial duration of the STIRAP pulse, bound-bound Rabi frequency
should be much larger than the free-bound one, and later it should
be just the opposite. Therefore, according to the arguments put
forward in Refs. \cite{javanainen2,quasicont}, the STIRAP
condition in a nondegenerate gas would not be satisfied and so
there is no STIRAP in a nondegenerate gas; though, there has been
a debate \cite{debate} on the possibility of STIRAP in a
nondegenerate gas. However, it is also argued \cite{javanainen2}
that, for Bose condensed atoms and in the thermodynamic limit, the
collective free-bound Rabi frequency becomes proportional to the
number density of atoms and so it regains a finite (nonzero)
value. Therefore, for condensed atoms, STIRAP becomes possible
because of the symmetric state bosonic stimulation
\cite{superchem}. In passing, we note that the possible effect of
a trap on nondegenerate free-bound-bound STIRAP was earlier
pointed out in Ref.\cite{quasicont}.

We emphasize that with trap-induced two-atom bound states, unlike
in the case two free atoms, two-photon Raman-type photoassociative
coherent dynamics is possible. The conditions for STIRAP can also
be satisfied for a sufficiently dilute nondegenerate atomic gas in
a harmonic trap due to the existence of trap-induced two-atom
bound states. However, the mean-field energy shift and inelastic
collisions such as the three-body interaction may limit the
efficiency of the STIRAP.

\section{results and discussions}

Having discussed in the preceding section the advantages we may
possibly derive from trap-induced atom-pair binding in performing
two-color photoassociative STIRAP in a nondegenerate atomic gas to
produce cold molecules, we now turn our attention to the effects
of this pair binding on  incoherent one-color photoassociative
processes such as spontaneous and stimulated linewidths. The
spontaneous linewidth for a transition $v \rightarrow n_t$ is
given by
\begin{eqnarray}
\gamma_{\rm sp}^{v\rightarrow n_t} =  \frac{4}{3\hbar c^3}
\omega_{vn_t}^3  |D_0|^2 |\eta_{v-n_t}|^2 + \gamma_v^{\rm
bound-bound},
 \label{bbsp}
\end{eqnarray}
where $\omega_{vn_t} = \omega_A- (\omega_v + \omega_{n_t})$,
$\omega_A$ is the frequency of the bare atom transition,
$\hbar\omega_v = E_v$ is the energy of the bound molecular level
$v$,  $\hbar \omega_{n_t}=E_{n_t}$ is the energy of the trap-bound
level $n_t$ and
\begin{eqnarray}
\gamma_v^{\rm bound-bound}\propto \sum_{v_g}\omega_{vv_g}^3
|\langle v |D(R)| v_g \rangle |^2
\end{eqnarray}
is the rate of spontaneous emission for transition from the
vibrational state $v$ in the excited molecular potential to the
bound levels in the ground molecular potential. $\omega_{vv_g}
=\omega_A-(\omega_v + \omega_{v_g})$, with $v_g$ as the
vibrational quantum number of the bound state $|v_g \rangle$ in
the electronic ground molecular potential.  For a bound-free
transition
\begin{eqnarray}
\gamma_{\rm sp}^{v\rightarrow \epsilon} =  \frac{4}{3\hbar
c^3}\int \omega_{v\epsilon}^3 |D_0|^2 |\eta_{v-\epsilon}|^2
d\epsilon + \gamma_v^{\rm bound-bound}
 \label{bfsp}
\end{eqnarray}
where $\omega_{v\epsilon} = \omega_A - (\omega_v +
\epsilon/\hbar)$. The integration on the right-hand side of the
above equation  is over the distribution of collision  energy
$\epsilon$.

The contribution to the total spontaneous line width due to
transitions from the ultralong-range bound states in the excited
molecular potential to bound states in the ground molecular
potential is negligible. Because the turning points of even the
least-bound state (longest range) in a typical ground molecular
potential that asymptotically becomes a van der Waals interaction
of the form $-C_6/R^6$ are  of the order of 1 nm, as can be
estimated from the binding energies of such states for Li$_2$ and
Na$_{2}$ as calculated in Refs. \cite{cote,psj}. In contrast, the
excited ultralong-range vibrational states, as calculated here by
the QDT method in Sec. IV, have their turning points at a much
larger distance of the order of 100 nm. We further note from the
plots of the wave functions of these ultralong-range vibrational
states (Fig. \ref{fig2}) that, below 10 nm, the amplitude of these
wave functions is vanishingly small. Therefore, the FC overlap
integral of these excited ultralong-range states with the bound
states in the ground electronic manifold would be negligible in
comparison to that with the two-atom scattering states as well as
trap-induced bound atom-pair states. We henceforth calculate the
spontaneous emission width for a transition from a particular
ultralong-range vibrational level in the excited molecular
potential only to the trap-induced bound atom-pair states and to
the two free-atom states (scattering states) while neglecting all
other molecular bound-bound transitions. In this context, a
pertinent question that may arise is whether such tight traps
would have any influence on the molecular bound states in the
ground molecular electronic manifold. As we have already
mentioned, the least-bound or the nearest-to-zero-energy molecular
bound states supported by  van der Waal's potential $-C_6/R^6$
have their outer turning point at a typical separation of 1 nm. If
we now estimate, for a typical value of $C_6 = 1500$ a.u., the van
der Waal potential energy at $R = 1$ nm, and compare this with the
relative harmonic trap energy of two atoms at the same separation
in an isotropic harmonic trap with harmonic frequency 100 kHz, we
find that this trapping potential energy is smaller than the van
der Waal potential by at least two orders of magnitude. So, for
harmonic trapping frequency $\omega_t \le 2\pi \times 100$ kHz,
the trap has negligible influence on the molecular bound states in
the ground electronic molecular potential. This is why in writing
the expressions for $\gamma_{sp}^{v\rightarrow n_t}$ and
$\gamma_{sp}^{v\rightarrow \epsilon}$ in Eqs. (\ref{bbsp}) and
(\ref{bfsp}), respectively, we have retained the same
$\gamma_v^{\rm bound-bound}$ in both expressions.

In Sec. III, we have described two-atom bound states
$\Psi_{n_t}(R)$ formed inside a harmonic trap.  Unlike the
free-bound case, the  FC factor for two trapped atoms can be
controlled by tuning the trap parameters. The lower the trap
frequency is, the longer the extension of the trap-induced
two-atom bound state. Thus, by lowering the trapping frequency,
the FC factor for the higher molecular vibrational states can be
increased. Therefore, the trap-induced bound state can facilitate
probing of ultralong-range molecular states. Figure \ref{fig1}
exhibits the lowest three $\Psi_{n_t}$ for two values of trapping
frequency $\omega_t=2\pi \times 100$ or $2\pi \times 10$ kHz.
Since $\xi_s \ll 1$, they deviate very little from the
harmonic-oscillator states. The bound-state energies and the outer
turning point of these states are given in Table \ref{tb1}.

The wave functions $\Psi_v(R)$ of the last seven vibrational
states from $v=33$ to $v=39$ calculated by the QDT method as
outlined in Sec. IV are plotted in Fig. \ref{fig2}. Their
vibrational energies $E_v$, outer turning points $R_t$, and the
internuclear distances $R_{\rm max}$ at which $|\Psi_{v}|^2$
attains maximum are given in Table \ref{tb2}. To evaluate $E_v$,
we employ Eq. (\ref{bs}) with a constant $K^0 = 0.623$. The
vibrational energies of these states have also been calculated
previously by a similar method \cite{bogao}. From Table \ref{tb2},
it is clear that the molecule in these states would spend most of
its time at an internuclear distance larger than 100 nm. In terms
of typical molecular scale, this length scale is enormous. Now, a
question naturally arises: is it possible to probe these states by
PA? To seek an answer to this question, we carry out an explicit
calculation of the FC factors in two different cases: (1) both
atoms being trapped in a single harmonic potential and (2) the two
atoms are in a free scattering state.

  \begin{figure}
  \includegraphics[width=3.25in]{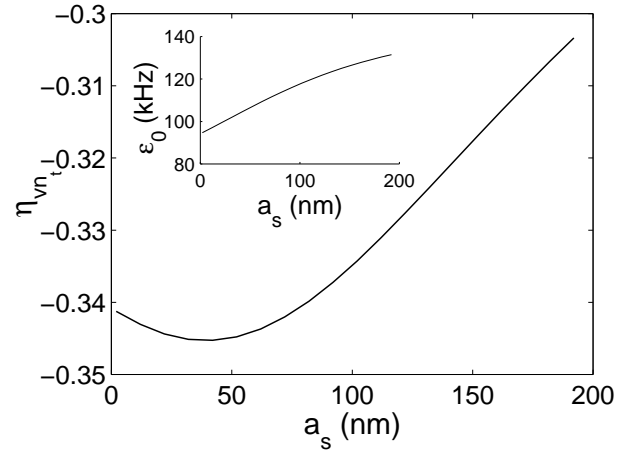}
 \caption{ Dimensionless   Franck-Condon
 integral $\eta_{vn_t}$ between molecular bound $v=35$ and the
 trap-bound $n_t=0$ (ground) states as a function of positive
 scattering length $a_{\rm sc}$ in nanometers for a
 trap frequency of $\omega_t=2\pi \times 10$ kHz.
 The inset shows the variation of trap-bound ground-state energy
 $\epsilon_0$ ($2\pi\nu_0$) as a function of $a_{\rm sc}$.}
\label{fig3}
 \end{figure}

 \begin{figure}
 \includegraphics[width=3.25in]{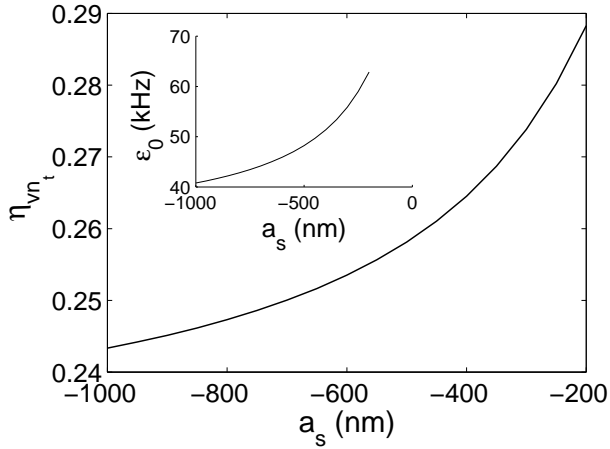}
 \caption{The same as in Fi.g \ref{fig3} but for a negative scattering length.}
\label{fig4}
 \end{figure}

In Table III, we  tabulate the values of the square of the FC
factor for selected states. Why and how much these values differ
from one pair of states to another can be naively understood from
the positions of their outer turning points and the shape of the
respective wave functions. For example, from Tables \ref{tb1} and
\ref{tb2}, we note that the states  $\Psi_{n_t=0}$ [for $\omega_t
= 2\pi\times 100$ kHz] and $\Psi_{v=33}$ have their outer turning
points at a comparable distance. Therefore, these two states
should have maximum overlap.

 \begin{table}
\caption{Square of Franck-Condon integral $|\eta_{vn_t}|^2$
between molecular bound $v$ and trap-bound states $n_t$ .}
\begin{tabular}{c c c c c c c}
&\multicolumn{3}{c}{$\omega_t = 2\pi\times10$ (kHz) }
 &\multicolumn{3}{c}{$\omega_t = 2\pi\times100$ (kHz) } \\
\multicolumn{1}{c}{$v$} & \multicolumn{1}{c}{$n_t=0$} &
\multicolumn{1}{c}{$n_t=1$} & \multicolumn{1}{c}{$n_t=2$} &
\multicolumn{1}{c}{$n_t=0$} & \multicolumn{1}{c}{$n_t=1$} &
\multicolumn{1}{c}{$n_t=2$}\\
 \hline
 33 & 0.075 & 0.074 & 0.059 & 0.221 & 0.122 & 0.009  \\
 34 & 0.102 & 0.080 & 0.039 & 0.024 & 0.066 & 0.019  \\
 35 & 0.116 & 0.004 & 0.033 & 0.000 & 0.214 & 0.021  \\
 36 & 0.009 & 0.022 & 0.000 & 0.000 & 0.013 & 0.093
 \end{tabular}
\label{tb3}
 \end{table}

In Figs. \ref{fig3} and \ref{fig4}, we display the variations of
the bound-bound FC factor as a function of the positive and
negative scattering length, respectively. Although as discussed
earlier that for a large positive $a_{\rm sc}$, energy-dependent
pseudopotential should be used in the calculation of trap-induced
atom-atom bound states, we simply took the energy-independent
pseudopotential for all the calculations. This approximation
underestimates the actual eigenenergies and hence underestimates
the actual range of those states. Therefore, for large positive
$a_{\rm sc}$, the actual value of the FC factor should be somewhat
larger. However, we took this approximation for our model
calculation and comparison because the overall pattern of
variation of the FC factor as a
 function of $a_{\rm sc}$ should remain the same.

When two atoms are not trapped, the FC factor becomes
 strikingly different. Since the FC
 factor in the two cases have different dimensionality, it is not
 easy to have a direct comparison between the two. However, we
 can make a comparison by comparing their spontaneous
 linewidth which is proportional to the respective FC factor.
 For the sake of this comparison with the bound-bound
 $v\rightarrow n_t$ transition, we have assumed that the the velocity
 of the free atoms follow a Maxwellian distribution at
 a temperature $T = E_{n_t}/k_B$
 in calculating the corresponding free-bound transitions, where
$k_B$ is the Boltzmann constant. The results are tabulated
 in Table \ref{tb4}. The free-bound FC factor of Eq. (\ref{fb}) depends critically on
the locations of the antinodes of the sinusoidal scattering state
of the two atoms relative to the outer turning point of the
molecular bound state. If the first antinode is located near the
turning point, then the free-bound FC factor would be greater than
if it is located far away from the turning point.

\begin{table}
\caption{Spontaneous linewidths $\gamma_{\rm sp}^{v\rightarrow
n_t}$ for transitions from  excited molecular  bound states  $v$
to trap-bound states $n_t$ are compared with corresponding
linewidth $\gamma_{\rm sp}^{v\rightarrow \epsilon}$ for molecular
bound-free transition. The frequency of the harmonic trap is
$\omega_{t} = 2\pi \times 100$ kHz.}
\begin{tabular}{c l l l l l l }
\multicolumn{7}{c}{$\gamma_{\rm sp}^{v\rightarrow n_t}/(2\pi)$ \&
$\gamma_{\rm sp}^{v\rightarrow \epsilon}/(2\pi)$
in (kHz), $k_BT = \hbar\omega_{n_t}$}\\
\hline
 &\multicolumn{2}{c}{$n_t = 0$} & \multicolumn{2}{c}{$n_t
= 1$} & \multicolumn{2}{c}{$n_t = 2$} \\
 \multicolumn{1}{c}{$v$}
&\multicolumn{1}{c}{$v\rightarrow n_t$}
 &\multicolumn{1}{c}{$v\rightarrow \epsilon $}
&\multicolumn{1}{c}{$v\rightarrow n_t$}
 &\multicolumn{1}{c}{$v\rightarrow \epsilon $}
&\multicolumn{1}{c}{$v\rightarrow n_t$}
 &\multicolumn{1}{c}{$v\rightarrow \epsilon $} \\
 \hline
 33 & 4278 & 28985 & 2883  & 40422 & 2359 &  47139  \\
 34 & 464 & 26591 & 5532  & 37216  & 1282 & 43467 \\
 35 & 0.002 & 23248 & 276   & 30781 & 4150 & 34831 \\
 36 & 0.011 & 6449 & 40.0  & 9434 & 245 & 11007 \\
 37 & 0.001 & 3429 & 6.3  & 4106 & 29.6 & 4249  \\
 38 & 0.000 & 599 & 0.5  & 623.7  & 2.4 & 735.2 \\
 39 & 0.000 & 10.2 & 0.002 & 7.8   & 0.009 & 7.4
 \end{tabular}
\label{tb4}
 \end{table}

 From Table
\ref{tb4}, we note that the tabulated bound-bound spontaneous
linewidths are much smaller than their respective free-bound
values. This suggests that if we PA two trapped atoms, and if the
resulting ultralong-range excited molecule remains trapped under
the same or a different trapping potential, then the spectral line
associated with the formation of the molecule may be resolved.

The trap frequencies we have considered for our numerical
illustration correspond to very tight traps.  In this section, so
far  we have restricted our treatment to an elementary
PA process of only two atoms inside an isotropic
harmonic trap. Now, the question obviously arises how the
   tight traps would affect the PA of an ensemble of
   atoms in a trap, in particular of an atomic BEC for its potential use
   in forming a molecular BEC by collective photoassociative
   STIRAP. In this regard,  it becomes necessary to consider
   various effects due to mean-field interactions and inelastic
   three-body recombination processes \cite{drummond2,becmolecule}.
   Let us first discuss the possible effect of a tight trap on
   the efficiency of STIRAP. The presence of a trap
   introduces an additional length scale in the system, namely, the size of the
   harmonic oscillator ground state $a_{t}$.
   For a typical magnetic trap in use today with Rb or Na atoms,
   trap frequency $\omega$ is of the order of $2\pi \times 100$ Hz.
   The frequencies we have considered here are 10-100 kHz,
   usually associated with strong far off resonant optical dipole traps.
   A condensate in such traps would typically have
   dimensions smaller than an order of magnitude of that in a magnetic trap. One
   can then argue that, for a typical number of atoms $N \sim 10^6$, the number density
   of atoms in such a tight trap would also be higher by  two to three orders of magnitude.
   Therefore, the strength of the mean-field interactions between atoms and also between
   atoms and STIRAP-produced molecules will be greatly enhanced. This increased
   mean-field interaction will lead to large
   frequency shifts and broadening in two-photon STIRAP PA spectrum, it also leads
   to a dramatically increased loss of atoms or molecules due to
   density dependent three-body collisions. Therefore, as already suggested
   in Ref. \cite{comment}, our study applies to cases of modest
   atom number densities associated with condensates of smaller numbers of atoms,
   of the order of 1000 or fewer.

   The enhancement of the Franck-Condon integral due to trap-induced binding between two
   ground-state atoms depends on the location of the outer turning point in the excited
   (S + P) molecular potential relative to the value of $a_{t}$. Our calculations
   reveal that the amplitude of the
   ground state of the two-atom (S-S) bound state  attains a maximum at about $a_{t}/2$.
   Therefore, the Franck-Condon integral will be maximal for transitions to states
   in the excited potential whose outer turning points lie close to $a_{t}/2$.
   For a typical magnetic trap of 100 Hz, $a_{t}$ is about 1 $\mu$m, while the outer
   turning point of the excited vibrational state used in Refs.
   \cite{drummond2,becmolecule} is at about 3 nm.
   The finding of Drummond {\it et al.} \cite{drummond2} that a weak trap reduces
   the efficiency of STIRAP, we believe therefore,
   is due to this huge mismatch between $a_{t}$ and the outer turning point.
   They have used the excited vibrational state whose
   free-bound transition frequency is 23 cm$^{-1}$ below the dissociation threshold.
   This state is far-off the dissociation threshold in comparison
   to the ultralong-range states we have described in the paper.
   To achieve a significant enhancement,
   the free-bound transition should involve the highest
   lying excited vibrational states, so that their outer turning points lie close to $a_{t}/2$.
   Normally, the excited ultra-long range vibrational states have their outer turning points
   at about tens to several hundred nanometers. In order to have large FC overlaps between these states
   and the trap-induced two-atom bound states, the trap frequencies should be much larger than
   100 Hz, i.e., we should use very tight traps.
   As long as the condition ($na_{\rm sc}^3 \ll 1$) remains fulfilled, the mean-field effects and
   three-body recombinations do not lead to substantial degradation of the efficiency of the
   STIRAP, as the gas remains sufficiently dilute and the interactions would be mainly of a
   two-body type, and only a small fraction of atoms will be pair correlated with trap-induced
   binding, they will experience enhanced FC coupling in a STIRAP. Presently,
   the nature of collision interaction between photoassociated excited or STIRAP-produced
   ground molecules and individual atoms is not precisely known.
   Therefore, we will omit a detailed study of how three-body
   recombination will affect the efficiency of the STIRAP. Intuitively it can be understood
   that with a high number density, the increased three-body recombination will lead to
   increased inelastic losses, which counteract
   any enhancement effect by tight trap.  However, it can also be
   intuitively argued that for a very low density of atoms and
   molecules, such three-body recombination losses may perhaps be overcome
   in achieving a considerably efficient STIRAP.

 In view of the foregoing discussion, it now appears that the best way
   out to avoid mean-field shifts and three-body effects is to
   photoassociate atoms
   in a Mott insulating state of atoms in an optical lattice \cite{mott}.
   In such a structured state of atoms, a pair of atoms can be arranged to occupy a
   single site of an optical lattice.

Before concluding this section, we would like to discuss another
important application of the ultralong-range excited vibrational
states in coherent laser spectroscopy of yet a different type of
potentially ultralong-range molecules, namely, molecules that can
be formed  in a controlled crossing of a Feshbach resonance
\cite{wieman}. These exotic molecules are produced in
magnetic-field-induced Feshbach resonance by  the nonadiabatic
coupling between two close molecular potential curves,  one of
which  acts as an energetically closed channel while the other as
an open asymptotic collision channel. Coherent transfer of atoms
into a molecular vibrational state close to the open-channel
threshold occurs through the tuning of a magnetic field. Recently,
these vibrational states have been numerically calculated
\cite{burnett} and shown to have an extent ranging from many tens
to many hundreds of nanometers. Although, these molecules are
translationally cold, they are not generally cold in their
vibrational states. Obviously, any two-photon resonant coherent
Raman spectroscopy of these molecules would involve higher-lying
excited vibrational states as we have calculated by the QDT
method. Thus, in addition to our finding of the trap enhancement
in coherent PA, our study of the QDT approach in finding
ultralong-range wave functions is important for applying the
STIRAP to the newly attained capability of a controlled crossing
of a Feshbach resonance, and the attempt to produce
translationally as well as vibrationally cold molecules.

\section{conclusions}

We have developed a scheme for estimating the FC factor involved
in a single-photon PA of two atoms into an ultralong-range
molecular state. In this approach, we have made two
simplifications: first, the initial state of the atoms are based
only on the asymptotic form at a large interatomic separation,
which in the low-energy limit, only depends on the s-wave
scattering length and second, for the excited molecular state
(corresponding to a S+P asymptote), we have used the exact
solutions of the asymptotic long-range molecular potential
$-1/R^3$, taking advantage of the QDT formulation. We have
calculated the wave functions of the molecular states that are
very close to dissociation threshold. We have calculated the FC
overlap integrals between these molecular states and the initial
states of the two ground-state atoms. We have shown that when two
atoms are trapped in a single harmonic potential well, the trap
frequency can significantly affect the integral and hence the PA
rate. We suggest that the ultralong-range excited molecule can be
formed by PA of two ground-state atoms inside a trap and
simultaneously switching on another trap that can confine
photoassociated molecules. In this way, the
 damping of the excited molecule can be minimized. We have also
 demonstrated the dependence of  the FC integral
 on both the sign and the magnitude of the atomic scattering length.
 To form a ground-state molecule by stimulated
Raman PA, the excited vibrational state should have sufficient
amplitude in the short-range region such that it can possess a
significant overlap with the ground molecular vibrational state.
For the model system studied in this paper, this seems practically
impossible. However, since trap potentials have a significant
effect on the overlap integral as well, it can perhaps also be
adjusted to affect stimulated Raman PA. In particular, when
performing coherent PA in a nondegenerate gas, we have shown that
a tight trap does offer an advantageous situation in comparison to
atoms in free-scattering state. However, as discussed in the
preceding section, the mean-field shifts and three body
recombination effects may ultimately limit the efficiency for
employing the STIRAP process to form a molecular BEC, and it
appears that the best way to exploit the trap-induced enhancement
of coherent coupling for producing a molecular BEC by STIRAP with
no adverse effects from three-body recombination and mean field
shifts is to photoassociate atoms in the Mott insulating state
with more than one atom per well of an optical lattice as first
investigated in Ref. \cite{mbecmott}. To this end, we would also
need an accurate method for calculating the ultralong-range
excited molecular vibrational states. Very recently,
far-off-resonance trap for atoms inside a cavity in the
strong-coupling regime  has been experimentally demonstrated
\cite{kimble}. This trap is insensitive to the internal atomic
states and has strong or tight trapping capability. Only a few
number of atoms can be captured in such a trap for a considerable
duration. Trap-induced enhancement may be experimentally studied
in  photoassociation inside such a tight trap without  any side
effects of mean-field interaction and three-body recombination.
Apart from the usual PA spectroscopy, our study can also find
applications in the future studies of coherent laser spectroscopy
of the recently discovered ultralong-range Feshbach molecules
\cite{wieman}.

\vspace{0.2cm}

\begin{center}
{\bf ACKNOWLEDGMENT}
\end{center}

Li You acknowledges the support from NSF.

 \end{document}